\documentclass{article} 
\usepackage{iclr2021_conference,times}
\usepackage[export]{adjustbox}

\usepackage{amsmath,amsfonts,bm}

\def\eqref#1{equation~\ref{#1}}

\def\1{\bm{1}}

\DeclareMathAlphabet{\mathsfit}{\encodingdefault}{\sfdefault}{m}{sl}
\SetMathAlphabet{\mathsfit}{bold}{\encodingdefault}{\sfdefault}{bx}{n}

\usepackage{hyperref}
\usepackage{url}
\usepackage{graphicx}
\hypersetup{
    colorlinks=true,
    allcolors=blue
}

\title{Inference of cell dynamics on perturbation data using adjoint sensitivity}

\author{Weiqi Ji$^{1}$, Bo Yuan$^{2,3}$, Ciyue Shen$^{2,3}$, Aviv Regev$^{3,4,5}$, Chris Sander$^{2,3}$, Sili Deng$^1$\\
$^1$ Department of Mechanical Engineering, Massachusetts Institute of Technology,\\ Cambridge, MA, 02139, USA \\

$^2$ Department of Cell Biology, Harvard Medical School, Boston, MA, 02115, USA \\

$^3$ Broad Institute of MIT and Harvard, Cambridge, MA 02142, USA \\

$^4$ Department of Biology, MIT, Cambridge, MA 02140, USA \\

$^5$ Current address: Genentech, 1 DNA Way, South San Francisco, CA, USA
}
\iclrfinalcopy 
\begin{document}

\maketitle

\begin{abstract}
Data-driven dynamic models of cell biology can be used to predict cell response to unseen perturbations. Recent work (CellBox) had demonstrated the derivation of interpretable models with explicit interaction terms, in which the parameters were optimized using machine learning techniques. While the previous work was tested only in a single biological setting, this work aims to extend the range of applicability of this model inference approach to a diversity of biological systems. Here we adapted CellBox in Julia differential programming and augmented the method with adjoint algorithms, which has recently been used in the context of neural ODEs. We trained the models using simulated data from both abstract and biology-inspired networks, which afford the ability to evaluate the recovery of the ground truth network structure. The resulting accuracy of prediction by these models is high both in terms of low error against data and excellent agreement with the network structure used for the simulated training data. While there is no analogous ground truth for real life biological systems, this work demonstrates the ability to construct and parameterize a considerable diversity of network models with high predictive ability. The expectation is that this kind of procedure can be used on real perturbation-response data to derive models applicable to diverse biological systems.
\end{abstract}

\section{Introduction}
Detailed information about interactions between molecules in biological processes can be an important ingredient for methods that aim to predict cellular behavior in response to perturbations and in deriving mechanistic models of disease processes \cite{pmid22383865}. Typical networks models in the biological literature are based on small scale targeted molecular experiments and while having led to a significant corpus of molecular biology knowledge, they are often incomplete and / or not quantitatively predictive. New modeling opportunities arise from modern perturbation screening and profiling techniques that have been used to generate rich biological datasets. These have greatly improved the potential for using machine learning technology to build much more comprehensive predictive interaction models according to what one might call the perturbation-interaction-response paradigm. 

Standard computational methods, however, are typically insufficient for accurate and efficient network inference. Static models, e.g., using partial correlation, maximum entropy or mutual information  do not use time-resolved information \cite{Locasale2009,Chan2017}. Dynamic models, e.g. dynamic Bayesian networks, and ordinary differential equation (ODE) models typically require prior knowledge of kinetic parameters, which are often not available \cite{M2005,hill2017}. Deep learning models, although usually providing good predictive ability, often lack interpretability to help understand the system \cite{zhou2015,Montavon2018}. Recent work of interpretable neural differential equations by \cite{rackauckas2020} combines the idea of neural ODEs \cite{neuralode} and physics-based models, providing a potential merge of deep learning and biological network inference for scientific machine learning models.

We have previously developed a hybrid approach, CellBox, that combines explicit ODE models of cellular interactions with an automatic differentiation framework implemented in TensorFlow (\cite{yuan2020cellbox} ). While successful for a specific cancer cell line, CellBox had some key limitations including, (1) we only tested in one specific system; (2) the model was trained only on steady-state data; (3) there  exists potential truncation errors during numerical ODE solving due to the use of non-stiff solvers on stiff biological systems, and (4) we used arbitrarily fixed time step, due to the fact that the computational graph is static in TensorFlow. 

Here, we implemented the CellBox algorithm in Julia  with the aim to move beyond these limitations. We replaced back propagation through time with adjoint sensitivity \cite{fabian2017} and changed  Heun's ODE solver with fixed time steps to high-order ODE solvers with adaptive time steps. Therefore, the new implementation would allow adjoint backward optimization on full trajectory simulation. Using simulated data from both abstract and biology-inspired networks, we demonstrate that the new implementation  can accurately predict the dynamic time course of cell response to unseen perturbations. Importantly, this method can efficiently infer molecular interaction parameters with high accuracy for various network structures. These results, taken together, strongly suggest the broad applicability of CellBox in a diversity of systems.

\begin{figure}[t]
\begin{center}
\includegraphics[width=1\linewidth]{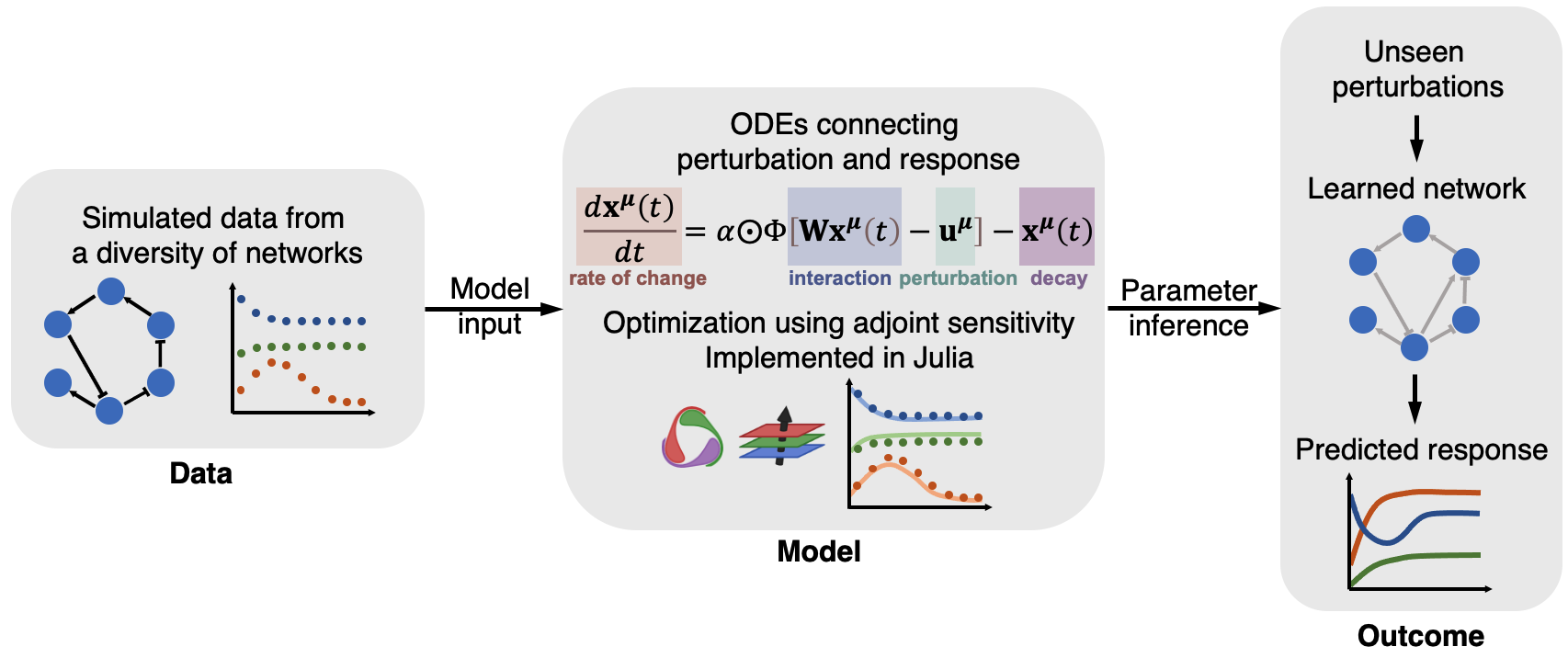}
\end{center}
\caption{Workflow for CellBox augmented by
adjoint sensitivity optimization.}
\end{figure}

\section{Method}
\label{method}

\subsection{CellBox model}

Following \cite{yuan2020cellbox}, a set of ODEs was used to describe the time development of the system variables upon perturbation (Eq. \ref{eqn2}).




\begin{equation}\label{eqn2}
\frac{d \mathbf{x}^\mu(t)}{d t} = \mathbf{\pmb{\alpha}} \odot \Phi \big[ \mathbf{W} \mathbf{x}^\mu (t) - \mathbf{u}^\mu \big] - \mathbf{x}^\mu (t)
\end{equation}

where the $n$ dimensional vector ${\mathbf{x}^\mu(t)}$ represents the change in molecular or phenotypic measurement, e.g., log ratios of molecular concentration after and before perturbation $\mathbf{u}$, and $n$ is the number of molecule types. The initial condition is $\mathbf{x}(0)=0$. $\mathbf{W} \in \mathbb{R}^{n \times n}$ quantifies the directional interaction among molecules where each ${w_{ij}}$ denotes the interaction from network node ${j}$ to node ${i}$. The scaling factor $\pmb{\alpha}$ quantifies the strength of the natural decay term $-\mathbf{x}^\mu(t)$, which is the tendency of an entity to return to its steady state level before perturbation. A hyperbolic tangent envelope function is used to introduce a saturation effect of the interaction term. $\odot$ denotes element wise multiplication.

By integrating Eq. \ref{eqn2} using an ODE solver, we can predict the time evolution of $\mathbf{x}^\mu(t)$ after a given perturbation $\mathbf{u}^\mu$.  CellBox learns these system parameters from experimental data.  With given input data, CellBox optimizes $\mathbf{W}$ and $\pmb{\alpha}$ by minimizing the mismatch between model predictions and experimental measurements. Here, instead of experimental data, we provide CellBox time-series pseudo-experimental data, generated from a given network model assumed to be the ground truth. Multiple time points are sampled and noise is added for different perturbation conditions, which collectively form the training data. Once trained, the models can predict the responses to unseen perturbations of the system.

\subsection{Adjoint sensitivity methods in dynamic system optimization}
\label{methods:node}

The dimension of $\mathbf{x}$ can be on the order of hundreds and thousands (human cells have about 20,000 genes in total), thus models can have  $10^4-10^8$ interaction parameters, leading to computational challenges for the parameter inference algorithm. Recently developed neural ODEs \cite{neuralode,rubanova2019latent} have shown promise in efficiently learning complex dynamic models from time-series data. One can view Eq. \ref{eqn2} as a neural network with single hidden layer where both the dimension of input and output is equal to $n$. We have re-implemented CellBox to learn the model parameters in the framework of neural ODEs. For instance, we denote the model predictions as

\begin{equation}\label{eqn3}
\hat{\mathbf{x}}^{\mu}(t) = ODESolve(\frac{d \mathbf{x}^\mu(t)}{d t};\mathbf{x}^\mu(t=0), \mathbf{W}, \pmb{\alpha})
\end{equation}

And we use the loss function of mean absolute error (MAE), i.e.,
\begin{equation}\label{eqn4}
Loss(\mathbf{W}, \mathbf{\alpha}) = \underset{\mu, t}{MAE}(\hat{\mathbf{x}}^{\mu}(t), \mathbf{x}^{\mu}(t))
\end{equation}

where $\hat{\mathbf{x}}^{\mu}(t)$ and $\mathbf{x}^{\mu}(t)$ correspond to simulated and measured molecular profiles. In this work, we use the ODE solver of the Tsitouras 5/4 Runge-Kutta method, implemented as Tsit5() in the Julia package of DifferentialEquations.jl developed by \cite{rackauckas2017differentialequations}. The first-order optimizer Adam proposed in \cite{kingma2014adam} implemented in Flux.jl by \cite{innes2018flux} is adopted with a default learning rate of 0.001. Learning rate annealing is exploited. We employ weight decay to  encourage sparsity in the interaction matrix (which is expected given our understanding of interactions in biological networks) and the value of weight decay is treated as a hyper-parameter. To accelerate the training, the training proceeds with a specific mini-batching, in which we sample $n_s=5$ time points from each perturbation condition as a single batch and we iteratively loop over all of the perturbation conditions. Unlike regular deep learning tasks, in addition to tracking the loss function, we also monitor the convergence of the model parameters by inspecting the Pearson correlations between learned interaction weights and ground truth network connections, reported as inference accuracy.

\begin{table}[ht!]
\textbf{\caption{Abstract synthetic systems}}
\begin{center}
\includegraphics[width=0.95\linewidth]{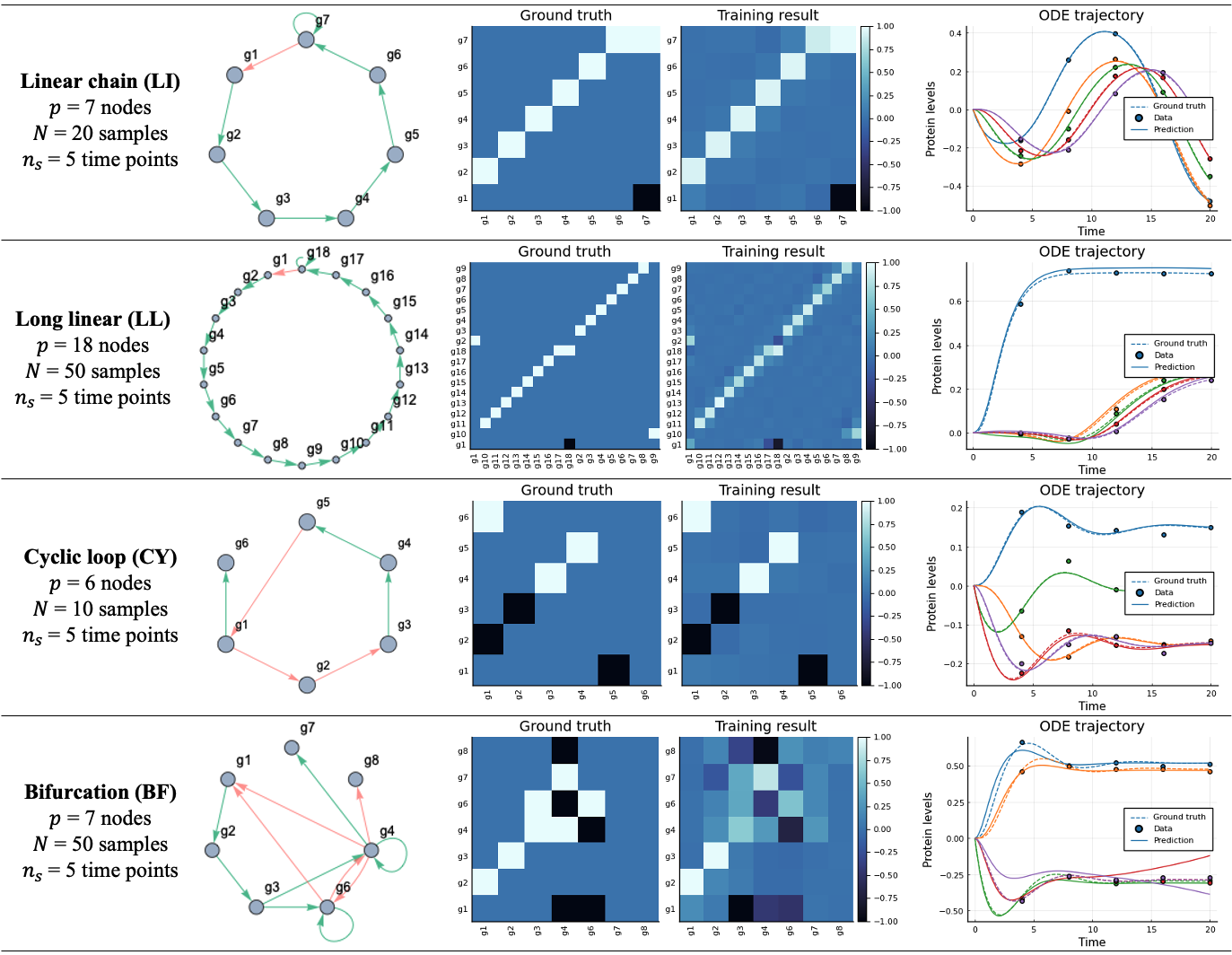}
\end{center}
\end{table}
\label{table:fig2}

\section{Results}
\label{results}

\subsection{Abstract synthetic  systems}

We first conduct proof-of-concept experiments by applying the CellBox algorithm to several typical abstract networks presented in \cite{beeline}, including linear (LI), cyclic (CY), and bifurcation (BI) networks. We artificially introduce perturbations into the systems and use ODE equations to simulate pseudo-experimental response trajectories, from which samples are drawn at multiple time points and used in training. To mimic realistic experimental setups, each synthetic perturbation condition is constrained such that no more than 2 nodes are perturbed simultaneously and at most 5 time points are sampled. We discuss the effect of the number of nodes perturbed and the number of time steps on model performance in Section 4.

We evaluated both the predictive accuracy of cell responses to these perturbations, as well as the accuracy of network inference. The model was trained with as few as 10 perturbation conditions (\textbf{Table 1}). Our results indicated that on these simulated  biological systems, the CellBox method could  train predictive models of system responses to external perturbations (average MAE $<10^{-3}$), as well as capture the system structures by inferring the network interaction parameters (average Pearson's correlation$ >0.9$). Overall, while different systems resulted in different efficiency of parameter inference, CellBox could be effectively trained for all four  system types tested, when provided with sufficient perturbation conditions. These results suggest the potential applicability of such methods to diverse biological systems.

\subsection{Biology-inspired synthetic systems}

\begin{table}[t!]
\textbf{\caption{Biology-inspired synthetic cellular systems}
}
\label{fig3}
\begin{center}
\includegraphics[width=0.95\linewidth]{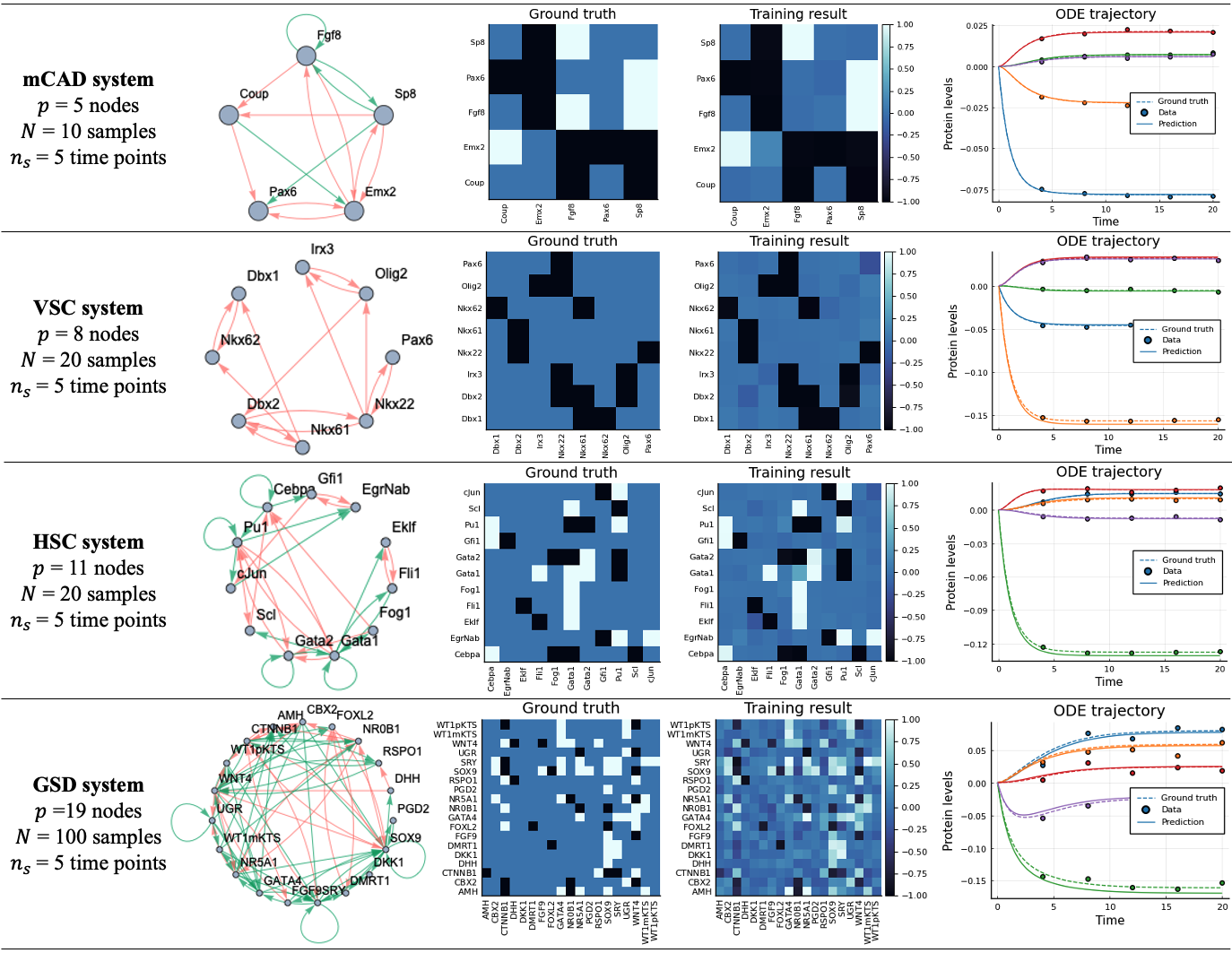}
\end{center}
\end{table}

We tested CellBox on synthetic analogs of four real biological systems adapted from \cite{beeline}, including mammalian cortical area development (mCAD), ventral spinal cord (VSC) development, hematopoietic stem cell (HSC) differentiation and gonadal sex determination (GSD), which were human-curated from the research literature and have been repeatedly validated by experiments. We then used these biology-inspired synthetic  system models in simulation and model training in the same way as the synthetic systems described above. Similar to the abstract systems, these models achieved the same level of prediction error (MAE: $10^{-4}$-$10^{-3}$) and inference accuracy (Pearson's correlation: $0.9$-$0.99$) (\textbf{Table 2}) and accurately predicted post-perturbational responses in each of the four systems described in \cite{beeline}. 

\section{Conclusion and discussion}
Constructing data-driven models for molecular interactions and cellular behavior of biological systems is useful but challenging.  Here we adapted the CellBox method \cite{yuan2020cellbox, Cellbox_NIPS}, a hybridization of explicit dynamic models and machine learning optimization, with adjoint sensitivity methods in Julia. We tested performance on a diverse range of synthetic network systems, and demonstrate that by simulating perturbation-response data and using such data to construct interaction models, CellBox methods can be effectively trained to predict responses and reconstruct the ground truth interaction network. The overall performance across diverse systems, including several typical abstract networks and multiple biology-inspired networks,  suggest that the CellBox can potentially be applied to a wide variety of biological systems.

Another improvement in this work compared to the first version of CellBox  is that the previous models used only single endpoint measurement, assuming the system had reached equilibrium.  It also obvious that time series data can be more informative for understanding and simulating the system, as explored in  (\cite{elin2020}). To  explore this potential we used simulated time series data for training. The  models then  are challenged to predict the full time series of the dynamic response to perturbations. This encouraged us to redesign the optimization procedure, e.g. by incorporating NeuralODE and adjoint sensitivity methods from \cite{neuralode, fabian2017}, as well as to challenge the model training efficiency and robustness, for which more careful fine-tuning are used to ensure stable simulation and modeling.

The data generation procedure used here does not systematically explore what are the minimal or optimal requirement of data for successful training. Some of the models above can be effectively trained with even less data, e.g., only 1 time point measurement; so it is not surprising that some biological datasets that are sparse in the time dimension are still informative. However, we did test scenarios where each synthetic condition has more than 2 nodes perturbed or more than 5 time points measured. As expected, increased information benefits model training and results in better performance even with a smaller sample size. Taken together, these observations might help design the scope of perturbation experiments needed to accurately simulate and predict the responses of a particular system. In future work, exploring associations of experiment size and quality with successful training would be of tremendous help in experimental design and validation. 







\bibliography{iclr2021_conference}
\bibliographystyle{iclr2021_conference}


\end{document}